\documentclass[pre,twocolumn,superscriptaddress,showpacs]{revtex4}
\usepackage{latexsym,amssymb,graphicx,amsmath}
\usepackage{epsfig}
\newcommand{\subscript}[1]{\ensuremath{_\textrm{#1}}}

\newcommand{\be}{\begin{equation}}
\newcommand{\ee}{\end{equation}}
\newcommand{\bea}{\begin{eqnarray}}
\newcommand{\eea}{\end{eqnarray}}

\def\figone#1#2#3{\begin{figure}
\centering \leavevmode
\epsfxsize=0.85\columnwidth \epsfbox{#1}
\caption{#2 \label{#3}}
\end{figure} }

\begin{document}

\title{Dynamics in the Metabasin Space of a Lennard-Jones Glass Former:  Connectivity and Transition Rates}

\author{Yasheng Yang and Bulbul Chakraborty}

\affiliation{Martin Fisher School of Physics, Brandeis University,
Mailstop 057, Waltham, Massachusetts 02454-9110, USA}

\begin{abstract}
Using simulations, we construct the effective dynamics in metabasin space  for a  Lennard-Jones glass-former.
Metabasins are identified via a scheme that measures transition rates between inherent structures,  and generates
clusters of  inherent structures by drawing in  branches that have the largest transition rates.  The effective
dynamics is shown to be Markovian but differs significantly from the simplest trap models.  We specifically show that
retaining information about the connectivity in metabasin space is crucial for reproducing  the slow dynamics observed
in this system.
\end{abstract}


\maketitle

\section{Introduction}
The origin of the dramatic slowing down of dynamics in supercooled liquids upon
approaching the glass transition temperature, has been of great
research interest. 
The potential energy landscape (PEL) has proved to be an important
conceptual tool for analyzing the dynamics of supercooled liquid
\cite{Goldstein1969,StillingerWeber1984,Stillinger1995,PhysRevE.60.6507,2000JChPh.112.9834S,Heuer0953-8984-20-37-373101}.
In PEL picture, the configuration hyper space
can be seperated into a collection of potential energy valleys,
each indentified with a local minimum or ``inherent structure". There have been many recent studies demonstrating that
the low temperatures dynamics of model supercooled liquids is dominated by activated dynamics 
between ``traps'' represented by metabasins which are clusters of inherent structures\cite{PhysRevLett.84.2168, PhysRevLett.91.235501,
2003PhRvE..67c0501D}.
In this activated regime, a natural mechanism for glassy dynamics is provided by the trap model\cite{1996JPhA...29.3847M}.  Extensive numerical simulations
of model glass formers have shown that the dynamics in metabasin space can be mapped on to the original\cite{PhysRevLett.90.025503} or an extended version of the trap
model\cite{2005PhRvE..72b1503H}.  The trap model framework neglects the connectivity in the metabasin space.
An interesting question to ask is whether a non-trivial connectivity in the metabasin space, such as the one found in
the multiple funnel landscape\cite{doye-1999-111, d.wales-2000},  leads 
to important differences from the trap model predictions regarding diffusion constants and relaxation timescales.   

An important ingredient in any framework for mapping the dynamics of a supercooled liquid to metabasin space is the method for constructing the metabasin.  In previous work,
metabasins were extracted from molecular dynamics trajectories projected on to the inherent structure space\cite{2005PhRvE..72b1503H}, where it was observed that there were long stretches
where the system transitioned back and forth between a finite number of inherent structures, and these were grouped into metabasins.   
The usefulness of the metabasin concept lies in the enormous simplification
in the dynamics that results from projecting the supercooled-liquid dynamics on to the metabasin space: this projected
dynamics is \cite{2005PhRvE..72b1503H}, and therefore, understanding glassy dynamics reduces to the 
problem of understanding
the properties of this random walk. Adopting the perspective that the aim of the metabasin construction is to define a space in which the complex dynamics of the supercooled liquid becomes Markovian,
opens up the possibility of constructing metabasins using other algorithms.   In this paper, we pursue this line of 
reasoning and use a metabasin construction scheme based on a knowledge of the transition rates
between different inherent structures.   With the metabasins defined, we  (a) test whether a trap model with these
metabasins
as the traps provides an adequate description of the observed dynamics of the supercooled liquid, (b) include
the connectivity in the metabasin space to extend the trap model predictions and show that the correlations extracted from this random
walk, which takes into account the non-trivial connectivity, 
agrees well with the ones measured directly in the molecular dynamics simulations,  and
finally (c) we analyze the properties of this random walk in order to gain some insight into the slow dynamics.  

\section{Model}
The model system we use to study the dynamics of supercooled 
liquid is a Lennard-Jones binary mixture (LJBM) consisting  of 67 particles,
53 of type A and 14 of type B.  The simulation box is a cube with periodical boundary condition.
All particles have the same mass $m$, and they interact via a Lennard-Jones potential
$v(r) =4\epsilon((\sigma/r)^{12}-(\sigma/r)^6)$,
where $r$ is the distance between them. 
The interaction parameters,
depending on the types of participating particles, are:
$\epsilon_{AA}=1.0$, $\sigma_{AA}=1.0$, $\epsilon_{AB}=1.5$, $\sigma_{AB}=0.8$,
$\epsilon_{BB}=0.5$ and $\sigma_{BB}=0.88$\cite{PhysRevLett.73.1376}.
In the following, all quantities will be expressed in reduced units, with the unit
of length as $\sigma_{AA}$, the unit of energy as $\epsilon_{AA}$, and the unit
of time as $(m\sigma^2_{AA}/\epsilon_{AA})^{1/2}$ \cite{AllenTildesley}.
The number density of particles is $1.2$. To accommodate the box with periodic boundary conditions, the potential is
shifted and truncated with a quadratic cutoff \cite{PhysRevA.8.1504,AllenTildesley,2003JPCM...15.1253S,2007JChPh.127l4504C},
which ensures continuity of the potential and its gradient at the cutoff radius.

The modified potential is
\begin{align}
v^c(r) = & v(r) - v(r_c) - \frac{(r^2-r_c^2)}{2r_c}\left(\frac{dv(r)}{dr}\right)_{r=r_c}\nonumber\\
 = & 4 \epsilon((\sigma/r)^{12}-(\sigma/r)^6\nonumber\\
 & +(6(\sigma/r_c)^{12}-3(\sigma/r_c)^6)(r/r_c)^2\nonumber\\
 & -7(\sigma/r_c)^{12} + 4 (\sigma/r_c)^6)
\end{align}
for $r<r_c$ and $v^c(r)=0$ for $r\geq r_c$.
Since the cutoff distance should be smaller than half of the box size,
a cutoff distance $r_c=1.9$ is used.
This cutoff makes the potential shallower than the original\cite{PhysRevLett.73.1376}, and, therefore, the ratio
between A and B particles is adjusted to minimize the chances of crystallization.
The velocity form of Verlet algorithm is used to solve the Newtonian equation, with $\delta t=0.001$.
The temperature range considered is from $T=0.7$ to $T=0.48$. In this regime
the supercooled liquid slows down significantly yet its equilibrium properties can
be studied in simulations.
Temperature is fixed by resampling the velocities from a Boltzmann distribution after every 10240 simulation steps.
Initial configurations are generated by first equilibrating the system  at $T=5$, then at $T=2$,
and then slowly 
cooling  down to the temperatures of interest 
with cooling rate $\leq 3.33\times 10^{-6}$ \cite{1998Natur.393..554S}
(Smaller cooling rates are used at lower temperatures).
At each of these temperatures, the system is 
equilibrated for a time much longer than (at least an order of magnitude) the estimated $\alpha$-relaxation time 
before any data is collected for measurement.

\section{Inherent Structures}
Inherent structures (IS) are obtained using conjugate gradient minimization techniques
during the simulation\cite{Stillinger1995,Heuer0953-8984-20-37-373101}.
Although, theoretically, inherent structures  can be labeled by their extact potential energy, there is a danger of
labeling two different
inherent structures with the same potential energy since any numerical procedures measure the potential energy with finite precision.
To alleviate this problem, we use a pair of energies to label each inherent
structure, $\{V,V_{BB}\}$,
where $V$ is the potential energy of local minima,
and $V_{BB}$ is the potential energy between type $B$ particles.
If two different inherent structures have the same $V$,
their $V_{BB}$ are likely to be different since they have different
arrangement of particles.  Implementing this procedure is most important for the metabasin cosntruction since many
thousands of IS are generated in the process.

Given an inherent structure, an ensemble of configurations in the same valley can be constructed by  using a
restricted Boltzmann sampling that is discussed in the next section. With each of these configurations, and
initial velocities sampled from the Boltzmann distribution, 
the waiting time out of this IS valley can be measured using interval-bisection method \cite{2003PhRvE..67c0501D}.
For the temperature range considered in this paper, 
it turns out that for many inherent structures,
the distribution of waiting time deviates from exponential significantly,
thus the transition between inherent structures is history-dependent.
The history dependence indicates that there is no clear separation of scales between the
time taken to equilibrate in an IS valley and the hopping between different IS.  This is reasonable given the fact
that there can be arbitrarily small dynamical barriers between the IS, which are defined solely based on their property
of being a potential energy minimum. 
It is expected that grouping IS into metabasins such that IS with frequent transitions between them are in
the same metabasin would lead to larger barriers between metabasins, and, therefore, a separation of time scales, and 
Markovian dynamics.

\section{Construction of Metabasins} Starting from one inherent structure in a hypothetical metabasin,
the system will more likely go into another inherent structure in the same metabasin.
To construct a metabasin from a randomly choosen inherent structure $A$,
we, therefor,  need to start the simulation from $A$ for many times,
and count how many times the system goes into each neighboring inherent structure.
There must be a most frequently visited neighbor, $B$.
Or in another word, $A$ has stronger connection to $B$, than to any of its other neighbors.
As shown in Figure~\ref{mbconstruction}, $A$ and $B$  get connected by an arrow.   The procedure is repeated starting
with $B$ and the process of building this cluster continues until we find a connection which links two IS that are
already in the cluster.  All the IS in the cluster are now assigned to one metabasin.
We then check the other neighboring IS found in the process of constructing the cluster and identify
their most frequently visited neighbor.  If these belong to the existing cluster, then these IS also get assigned to
the  metabasin containg the starting configuration, $A$.
\figone{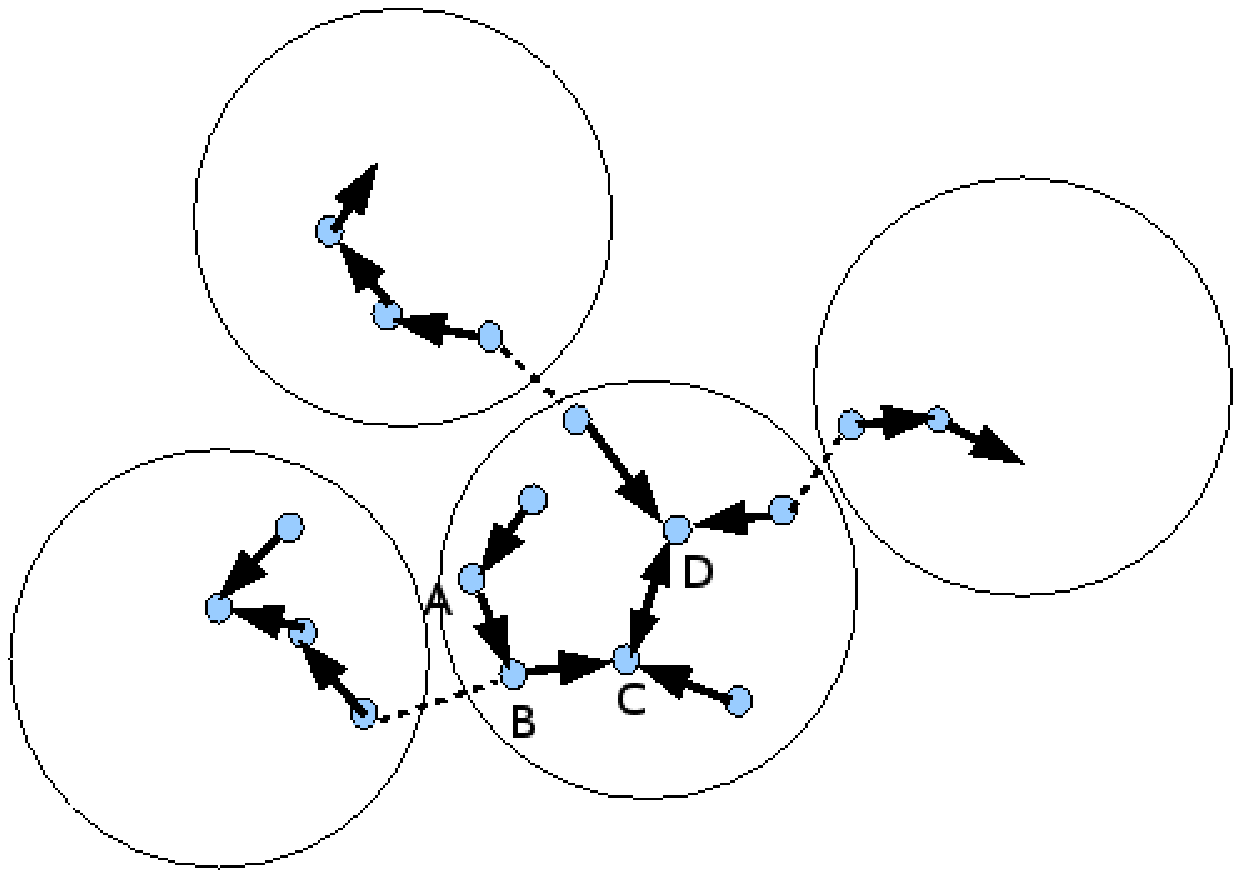} {Illustration of the construction of metabasins.
Small circles denote inherent structures.
Solid arrows denote the strongest transition branches from corresponding metabasins.
Each big circle encloses a cluster of inherent structures, i.e., a metabasin.
Dash lines denote the weaker inter-metabasin transitions}{mbconstruction}
During the construction of metabasin, tens of thousands of inherent structures
are encountered, which crowd into a small range of potential energy, and as mentioned earlier, it becomes essential to
label them with both $\{V,V_{BB}\}$.  

Our construction is similiar to the hierarchical master equation approach in \cite{mauro:184511}.
Comparing with MB construction methods based on trajectories, our approach is objective 
except for the fact that the computational force is limited to explore all the details of a MB in the phase space.

The construction of the metabasin, described above, relies on the transitions
between IS being history independent.  It is, therefore, important to obtain an initial {\it ensemble} of states which are
in equilibrium in the  valley $A$ and use these as starting configurations to measure the number of transitions
to other IS.  All of these starting configurations  should have the same set of values for $\{V,V_{BB}\}$ but a
different set of velocities.  We
implement this restricted Boltzmann sampling by 
starting from an initial configuration $A_0$,  with initial velocities sampled from the Boltzmann distribution. and run
the simulation for $n$ MD steps. 
If the final state $A_1$ still belongs to the same inherent structure,
then we accept it as a member of the ensemble.  If not, we count $A_0$ again as a 
member of the ensemble.
This process is repeated until we
have a large enough ensemble of initial states to measure the transitions from.
In practice, $n$ is chosen to be $1/4$ of the estimated waiting time of the given IS,
and in order to reduce correlation,  the above process is repeated $8$ times before the resulting state is accepted as
initial state. The number of initial states is at least 40, and more often chosen to be greater for better estimation of
branching ratios. 

The temperature chosen to construct metabasin is $T=0.52$, a temperature  where IS are well defined and the dynamics is not prohibitively slow for metabasin construction.
The number of initial inherent structures, randomly chosen from MD trajectories is 140 and each of these constitute
the starting point  for metabasin construction according to the algorithm described above s.

The potential energy, $E $, of a metabasin is defined to be the energy of the most probable inherent structure in
it.

\section{Markovian property}A hallmark of Markovian processes is that the distribution of waiting
times is exponential\cite{Van-Kampen}.  Since the motivation behind the construction of the metabasins
is obtaining a Markovian model of the dynamics of supercooled liquids, the first task is to check whether our
metabasin construction yields a space in which the dynamics is Markovian.
To measure the waiting times,
an ensemble of initial states is constructed through a restricted Boltzmann sampling
in each metabasin and at the temperature of interest.  A constant temperature trajectory is then started from
each initial state, and interval-bisection method \cite{2003PhRvE..67c0501D} is used to determine
the inherent structure sequence of the trajecotry.
If the inherent structures of two succesive quenchs are not in the original
metbasin, the system is considered to have made a transition to a new metabasin, and
the waiting time is recorded. Since the conjugate gradient method used
to minimize the potential energy occasionally leads to a wrong
inherent structure, two succesive quenchs are used to signal a transition.
For all the metabasins constructed according to our algorithm, the waiting time 
distributions are founded to be exponential and, therefore, the dynamics in this space is indeed Markovian.
The mean waiting time $\tau_i$, for a metabasin $i$ with energy $E_{i} $
is measured in the temperature range $0.44 \leq T \leq  0.70$. 
This mean waiting time is found to be above the ballistic time region, indicating  the separation of time scales that we
would like to see for metabasins.
The mean waiting time is characteristic
of an activated process 
with:
\be
\label{eqfittauT}
\tau_i(T) \approx \tau_i^\infty \exp(E_i^{act}/T)~.
\ee
The effective activation energy $E^{act}$ is not strongly correlated with the energy $E $ although $E^{act}$
tends to be higher for lower lying metabasins, as shown in Figure~\ref{Eact_EMB}.
\figone{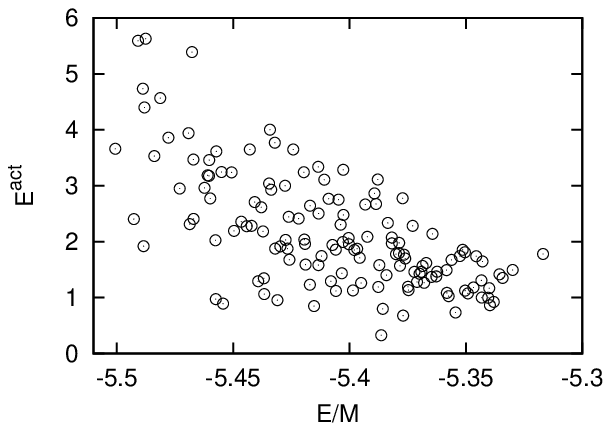} {Effective activation energy $E^{act}$ of metabasins,
plotted against their energy per particle, $E/M$, where $M$ is the number of particles. 
The activation energies are calculated according to Eq.~\eqref{eqfittauT}.}{Eact_EMB}
The time scale, $\tau(T)$ demonstrates a stronger correlation with a marked increase for deeper metabasins, as shown in
Figure~\ref{tauMB}. Since the dynamics in metabasin space is Markovian,  the evolution of the probability distribution
of metabasins, $P_{i}(t)$, is described by:
\be
\frac{d}{dt}P_i(t) = - \frac{1}{\tau_i(T)}P_i(t) + \sum_j w_{ij} P_j(t)
\label{eqrwalk}
\ee
where $\tau_i$ is the mean waiting time of metabasin $i$ at temperature $T$, and $w_{ij}$ is  the
transition rate from metabasin $j$ to metabasin $i$.
Envisioning the metabasins as traps of depth $E_{i}^{act}$, the simplest trap model is one that ignores the
connectivity and characterizes the trap space by $\tau_{i}$ and a density of states\cite{1996JPhA...29.3847M}:
\be
\frac{d}{dt}P_i(t) = - \frac{1}{\tau_i(T)}P_i(t) + \omega(t)\rho(E_{i}^{act})
\label{eqtrap}
\ee
where $\omega(t)$ is a normalization factor and $\rho(E^{act})$ is the density of traps with a depth $E^{act}$.
In most mappings of metabasin dynamics, in Lennard-Jones systems,  to trap models the energy of the metabasin,
$E$ has been considered as the depth of the trap.  As discussed earlier, we find that the activation energy and the
nominal energy of a metabasin are not strongly correlated.   We, therefore, choose to construct an effective trap
model that describes dynamics in an energy landscape, by 
averaging $\tau_{i}$ over metabasins with energies in a small range around a value $E$:
\begin{align}
\frac{d}{dt}P(E,t) = - \frac{P(E,t)}{\tau(E,T)} + \rho(E,T)\omega(t)
\end{align}
where
\begin{align}
\label{PET}
P(E,t)&=\sum \delta(E_i-E)P_i(t) \nonumber\\
\frac{1}{\tau(E,T)}&=\frac{\sum_i\delta(E_i-E)P_{i,eq}(T)/\tau_i(T)}{\sum_i\delta(E_i-E)P_{i,eq}(T)}\nonumber\\
\rho(E,T)&=P_{eq}(E,T)/\tau(E,T)\nonumber\\
\omega(t) &= \frac{\sum_i P_{i}(T)/\tau_i}{\sum_i P_{i,eq}(T)/\tau_i},
\end{align}
with $P_{i,eq}(T) \propto \exp(-\beta E_{i})$, the equilibrium probability of finding metabasin $i$, and
$P_{eq}(E,T)$ the equilibrium distribution of metabasin energy, which can be measured in the simulation. 
As shown below, this effective trap model differs from the simple
trap model\cite{1996JPhA...29.3847M} in that $\rho(E,T)$  depends on
temperature; reflecting the fact that $E_{i} $ is not the trap depth of metabasin $i$.

The metabasin energy distribution $P_{eq}(E,T)$ is obtained from simulations run at $T=0.6$, $0.52$ and $0.48$, and is found to be well
described by a Gaussian over this range\cite{PhysRevLett.83.3214,PhysRevE.60.6507,PhysRevLett.90.025503,PhysRevLett.91.235501}.
We have extracted the density of states from these distribution and find that $\Omega(E) \propto P_{eq}(E,T) \exp(E/T)$ is,
indeed, independent of temperature as show in Figure~\ref{rhoMBBoltzmann}. 
We use this observation to construct $P_{eq}(E,T)$ at temperatures other than the ones where it was measured explicitly,
and combine this information with $\tau(E,T)$ to obtain $\rho(E,T)$.   This function,  shown in Figure ~\ref{rhoMB},
depends on temperature and is different from $\Omega(E)$.  

The effective  trap model based on the deduced forms of $\rho(E,T)$, $\tau(E,T)$ and $\omega(t))$, provides a
description of the activated dynamics near the glass transition.  It is, however, a meanfield model that ignores the
connectivity in the
metabasin space.  In order to evaluate the effects  of the connectivity on the dynamics, we have calculated various
correlation functions using this trap model and compared it to the results of the direct, molecular dynamics
simulations.

\section{Testing the Trap Model} In this section, we compare the predictions of the effective trap model to
actual simulation results.
In the trap model, correlation functions are calculated by averaging over the sampled  metabasins according to:
\begin{align}
\label{eqexpect}
\langle A\rangle_{eq}&=\sum_i P_{i,eq}(T)A_i\nonumber\\
&=\int dE P_{eq}(E,T)\frac{\sum_i P_{i,eq}(T)\delta(E-E_i) A_i}{\sum_i P_{i,eq}(T)\delta(E-E_i) } \nonumber\\
&=\int dE P_{eq}(E,T)\frac{\sum_\alpha \delta(E-E_\alpha) A_\alpha}{\sum_\alpha \delta(E-E_\alpha)}
\end{align}
Here $A$ is the physical
observable of interest.
The first summation,  $\sum_i$ is over all metabasins, while $\sum_\alpha$ is over the metabasins, sampled according to
the Boltzman distribution in the simulations.   As mentioned earlier,  a total of 140 metabasins were sampled.  
In numerical calculation, the $\delta$ function in \eqref{PET}, is replaced by a
Gaussian with a narrow width, $2\sigma^2=0.01^2M^2$, with $M=67$ being the number of particles.

\figone{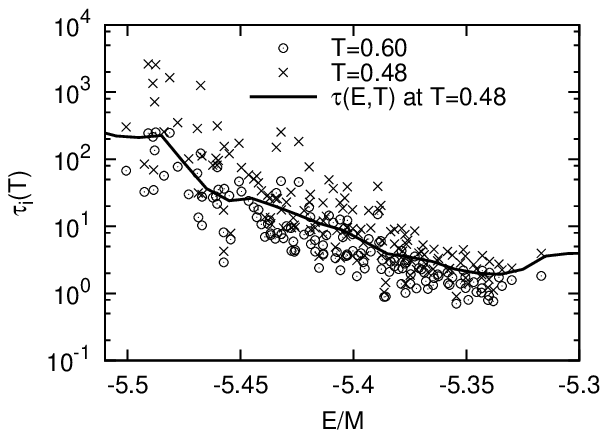} {Waiting time of metabasins measured at $T=0.6$ and $0.48$,
plotted against their metabasin energy per particle. The solid line is $\tau(E,T)$ calculated
according to Eq.~\eqref{PET}}{tauMB}
\figone{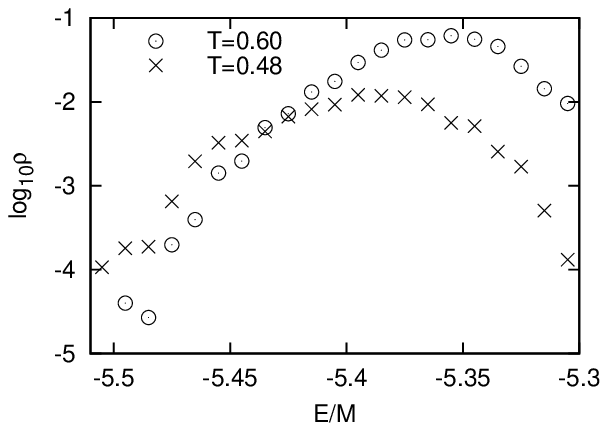} {$\rho(E,T)$ at  $T=0.6$ and $0.48$,
calculated according to Eq.~\eqref{PET}}{rhoMB}
\figone{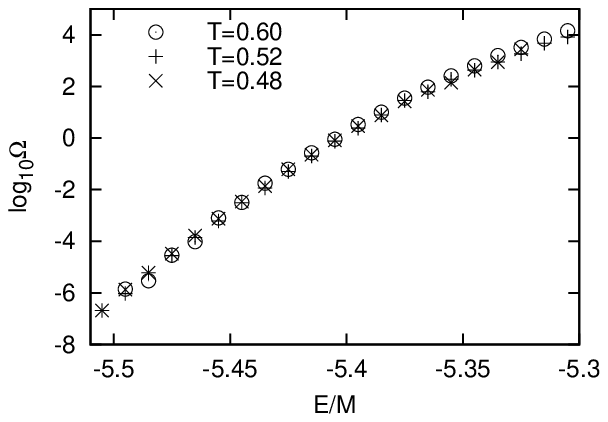} {Plot of the density of states of metabasins, $\Omega(E)$, shifted vertically to make
lines overlap.}{rhoMBBoltzmann}

\subsection{Estimate of the alpha relaxation time}
The $\alpha$ relaxation time $\tau_\alpha$ was measured in the simulation through its usual definition: $F_s(q_{0},\tau_\alpha)=1/e$, where $F_s(q,t)$ is the incoherent scattering
function\cite{hansen86a}.
In this paper, $F_s(q,t)$ is measured for type $A$ particles and  $q_{0} = 7.251$ \cite{PhysRevLett.73.1376}.
The $\alpha$ relaxation time can also be estimated from the distribution of single particle
displacements\cite{2006PhRvE..73a1504S}, which shows marked non-Gaussianity at timescales of the order of 
$\tau_\alpha$.  The non-Gaussianity is related to caging and a caging time scale, $\tau_{s}$ can be extracted by
measuring the time at which the probability of a type $A$  particle having a translation less than $1/q_{0}$
diminishes to $1/e$.   As shown in Table~\ref{TABalpha}, in the temperature range examined, $\tau_{s}$  and $\tau_{\alpha}$
are proportional to each other.  The caged motion of particles has also been related to hopping between
metabasins\cite{2003PhRvE..67c0501D}.

In the trap model framework, the $\alpha$ relaxation process corresponds to hopping between
metabasins\cite{2003PhRvE..67c0501D,PhysRevLett.90.025503}, and in this regime, the  correlation function $C(t)$ can
be constructed by assuming that 
the correlation is unity when the system is in the same metbasin and drops to zero as it leaves the
metabasin\cite{1996JPhA...29.3847M}.  For our effective trap model, this approximation leads to:
\be
\label{trapcorr}
C(t) = \langle\exp(-t/\tau_i)\rangle_{eq}
\ee
The expectation value is calculated according to Eq.~\eqref{eqexpect}.
In Figure~\ref{CT},  $C(t)$ is plotted along with $F_s(q,t)$.  As expected the shapes of the two functions are
different since $C(t)$ has no information about the short time dynamics.   For $C(t)$ to be a
useful  tool for understanding the glass transition, however, the change in time scale of the long-time dynamics should
closely resemble that of $\tau_{\alpha}$.
Within the trap model, a measure of the $\alpha$ relaxation time is provided by the  relaxation time $\tau_c$ defined by $C(\tau_c)=1/e$, and measured
using Eq.~\eqref{trapcorr}. The values of $\tau_c$ along with the values of $\tau_\alpha$
are listed in Table~\ref{TABalpha} for $T=0.6$, $0.52$, and $0.48$ and plotted 
in Figure~\ref{timescale}.
It is clear  that $\tau_c$ is much smaller than $\tau_\alpha$ for all temperatures listed, and more importantly, 
$\tau_\alpha$ increases significantly faster than $\tau_c$.
These results indicate that the trap model, as defined up to now, does not capture all of the physical processes
leading to the slow dynamics.

\figone{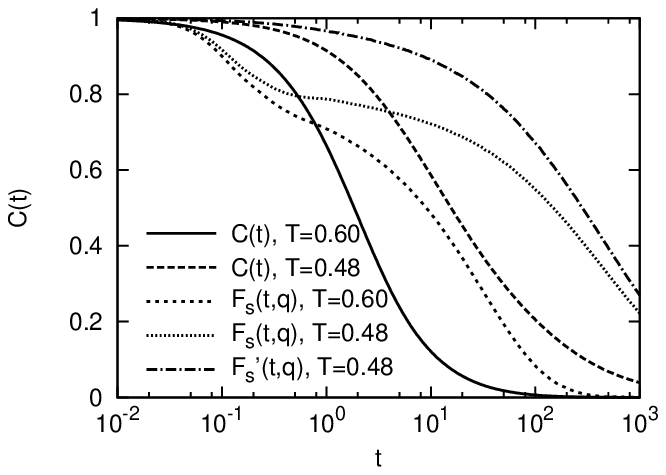} {Comparison of $C(t)$ obtained from the effective trap model at $T=0.6$ and $0.48$ and the incoherent
scattering function, $F_s(q,t)$ measured in simulations.  Also shown is $F_s'(q,t)$, the incoherent scattering function
measured from the inherent structure coordinates.  Time is measured in scaled units ({\it cf} text).}{CT}

\begin{table}
\caption{Comparison of  estimates of the $\alpha$ relaxation time, $\tau_\alpha$, $\tau_s$, and $\tau_c$.}
\label{TABalpha}
\begin{tabular}{c|cccccc}
\hline\hline
T &
$\tau_\alpha(T)$ &
$\tau_s(T)$ &
$\tau_c(T)$ & 
$\frac{\tau_\alpha(T)}{\tau_\alpha(0.6)}$ &
$\frac{\tau_s(T)}{\tau_s(0.6)}$ &
$\frac{\tau_c(T)}{\tau_c(0.6)}$ \\
\hline
0.9 &2.2 &6.7 & &0.11 &0.11 &\\
0.6 &20 &62 &2.9 &1 &1 &1\\
0.52 &91 &$2.7\times 10^2$ &1.1 &4.5 &4.4 &3.8\\
0.48 &$3.7\times 10^2$ &$1.1\times 10^3$ &32 &19 &18 &11\\
0.46 &$8.4\times 10^2$ &$2.5\times 10^3$ & &42 &40 &\\
\hline\hline
\end{tabular}
\end{table}

\figone{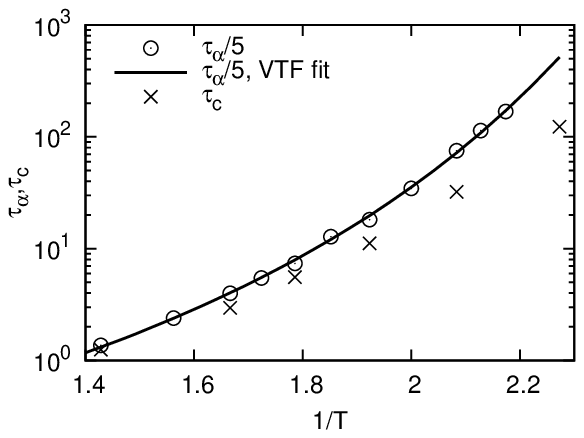} {Temperature dependence of $\tau_\alpha$ and $\tau_c$, measured in scaled units.
The solid line is the fit of $\tau_\alpha$ to the Vogel-Tamman-Fulcher form.}{timescale}

\subsection{Trap model prediction of the mean squared displacement (MSD)}
The MSD $\langle\delta r^2(t)\rangle$ of type A particles is measured directly from the simulations,  and as shown
in Figure~\ref{MSD},  there is caging.   The diffusion constant can be measure by looking at the long-time behavior:
$D=\langle\delta r^2(t)\rangle/6t$ at large $\langle\delta r^2(t)\rangle$.   The temperature dependence of $D$
is shown in Figure~\ref{diffusionconstant}.

\figone{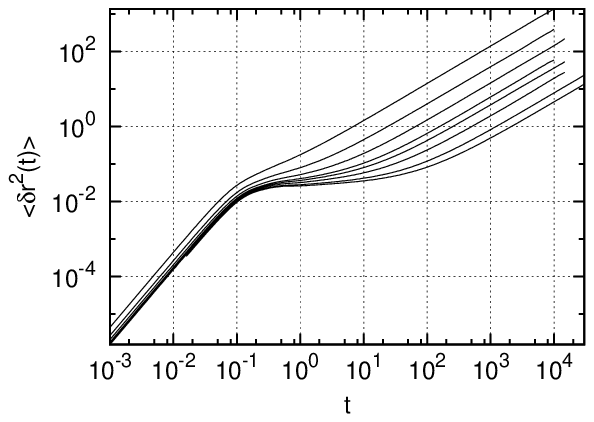}{MSD of type A particles plotted as a function of time $t$ measured in scaled units. The lines correspond to
$T=1.5$, $0.9$, $0.7$, $0.6$, $0.56$, $0.52$, $0.48$, and $0.46$, top to bottom at $t \ge 1$.}{MSD}

The MSD of particles in space can be mapped on to a random walk between metabasins \cite{2001JChPh.11410402S,2003PhRvE..67c0501D}.
Consider a random walk going through a sequence of $N$ metabasins,
$\alpha_1 \rightarrow\alpha_2 \rightarrow ... \rightarrow \alpha_{N+1}$,
where $\alpha_n$ is the label of metabasin.
The metabasin coordinate sequence is labeled as
$\vec \xi_{1} \rightarrow \vec \xi_{2} \rightarrow ... \rightarrow \vec \xi_{N+1}$.
where $\vec \xi_n$ is a $3M-component$ vector representing the coordinates of $M$ particles in metabasin $n$. 
The diffusion constant, $D_{MB}$,  of this random walk can be measured from the large $N$ limit:
\begin{align}
6M D_{MB}=\frac{\langle(\vec \xi_{N+1} - \vec \xi_1)^2\rangle_{rw}}{N\langle\tau_n\rangle_{rw}}~,
\end{align}
where $\langle\tau_n\rangle_{rw}$ is the average waiting time obtained by averaging over all random walk trajectories in the
metabasin space.
For each metabasin $\alpha_n$, the probability of being at metabasin $i$ is proportional
to $P_{i,eq}/\tau_i$. 
Thus $\langle\tau_n\rangle_{rw}$ can be calculated from $\langle 1/\tau_i\rangle_{eq}$,
which is an average over all metabasins weighted with equilibrium probabilities,
\begin{align}
\langle\tau_n\rangle_{rw}= \frac{\sum_i \frac{P_{i,eq}}{\tau_i}\tau_i}{\sum_i \frac{P_{i,eq}}{\tau_i}} = \frac{1}{\langle 1/\tau_i\rangle_{eq}}
\end{align}
Assuming that at each metabasin transition the displacement in $\vec \xi$-space is $\sqrt{\Delta}$, the diffusion
constant $D_{MB}$ is, therefore, $D_{{MB}} =
(\Delta /6 M) \langle 1/\tau_i\rangle_{eq}$. As shown in Fig.
\ref{diffusionconstant},  $D_{{MB}}$ rides above the $D$ measured from the MSD of particles, again, demonstrating that
the trap model picture, which ignores all connectivity in the metabasin space, underestimates the slowing down of the
dynamics.

\section{Effects of Connectivity in Metabasin Space}

One feature of the metabasin space that has been ignored in the trap model is that the Markovian dynamics describes a
random walk on a graph with non-trivial connectivity.   The calculation of the diffusion constant, $D_{MB}$ can be
improved upon by incorporating aspects of the connectivity that are captured by the transition rates $w_{{ij}}$.
Defining, $\delta \vec \xi_n = \vec \xi_{n+1} - \vec \xi_{n}$, the mean squared displacement in $\vec \xi$-space can be
written as:
\begin{align}
\label{msdfull}
\langle&(\vec \xi_{N+1} - \vec \xi_{1})^2\rangle_{rw}
= \langle(\sum_{n=1}^N \delta \vec \xi_n)^2\rangle_{rw} \nonumber\\
&= \sum_{n=1}^N\langle(\delta \vec \xi_n)^2\rangle_{rw} + 2\sum_{n=2}^N\langle\delta \vec \xi_{n-1} \cdot \delta \vec \xi_n\rangle_{rw} \nonumber\\
&+2\sum_{m=2}^{N-1}\sum_{n=m+1}^N\langle\delta \vec \xi_{n-m} \cdot \delta \vec \xi_n\rangle_{rw}\nonumber\\
&\approx N (\Delta + 2\langle\delta \vec \xi_{n-1} \cdot \delta \vec \xi_n\rangle_{rw})
\end{align}
Here we ignored $\langle\delta \vec \xi_{n-m} \cdot \delta \vec \xi_n\rangle_{rw}$ for $m>1$.
Then the expression of $D_{MB}$ becomes
\begin{align}
6M D_{MB}&\approx \langle 1/\tau_i\rangle_{eq}(\Delta + 2 \langle\delta \vec \xi_{n-1} \cdot \delta \vec \xi_n\rangle_{rw})
\end{align}

To calculate $\langle\delta \vec \xi_{n-1} \cdot \delta \vec \xi_n\rangle_{rw}$, 
one first picks the metabasin $\alpha_n$  at time step $n$ to be a specific one, say $i$, and 
averages over all possible metabasins $j$ at step $n-1$,
and over all possible metabasins $j'$ at step $n+1$, followed by 
averaging  over all possible metabasins $i$ at step $n$.
For a particular $\alpha_n=i$,
the probability of having a metabasin $j'$  at $n+1$  is $w_{j'i}/\sum_k w_{ki} = w_{j'i}\tau_i$.
Similarly,
the probability of having a metabasin $j$ at step $n-1$ must be $w_{ji}\tau_i$.
Thus
\begin{align}
\langle\delta \vec \xi_{n-1} \cdot \delta \vec \xi_n\rangle_i 
&= \langle\delta \vec \xi_{n-1}\rangle_i \cdot \langle\delta \vec \xi_n\rangle_i\nonumber\\
&=(\sum_j w_{ji}\tau_i\delta\vec \xi_{ij}) \cdot (\sum_{j'}w_{j'i}\tau_i \delta\vec \xi_{j'i})
\end{align}
with $\delta\vec \xi_{ij}\equiv \vec \xi_{i} - \vec \xi_{j}$.
Two facts should be noticed, (1) the dimension of $\vec \xi$, i.e.,
the number of independent coordinates,
$d=3M-3=198$, is very high,
and (2) For many metabasins the connectivity is sparse, with strong connections only to one other metabasin.   An
example is provided by the metabasins A
and B in Table~\ref{TABbranch}.  Because of (1),  any pair $\delta\vec \xi_{ij}$ and $\delta\vec \xi_{j'i}$
with $j'\neq j$ can be considered as two random vectors in a  high dimensional space, and
hence,  $\delta\vec \xi_{ij} \cdot \delta\vec \xi_{j'i}\approx 0$.
The exception is of course when $j=j'$, {\it i.e.}, when the system hops back and forth between two metabasins $i$ and
$j$, and 
$\delta\vec \xi_{ij} \cdot \delta\vec \xi_{j'i} = - (\delta\vec \xi_{ji}) ^2 = -\Delta$.
So we have
\begin{align}
\label{xidotxi}
\langle\delta \vec \xi_{n-1} \cdot \delta \vec \xi_n\rangle_i &\approx - \sum_j (w_{ji}\tau_i)^2 \Delta\nonumber\\
&= - p_i^2 \Delta
\end{align}
where $p_i^2\equiv\sum_j (w_{ji}\tau_i)^2$ is the probability 
of return to the previous metabasin.
For those metabasins that have many neighbors,
each with similiar (and low) branching ratio $w_{j'i}\tau_i$,
such as metabasin C and D in Table~\ref{TABbranch},  
the return probability is small.

Averaging Eq.~\eqref{xidotxi} over all possible metabasins $i$, one obtains, 
\begin{align}
\langle\delta \vec\xi_{n-1} \cdot \delta \vec \xi_n\rangle_{rw}
&= \frac{\sum_i \frac{P_{i,eq}}{\tau_i} \langle\delta \vec \xi_{n-1} \cdot \delta \vec \xi_n\rangle_i}{\sum_i \frac{P_{i,eq}}{\tau_i}}\nonumber\\
&\approx -\frac{\Delta \langle p_i^2 /\tau_i\rangle_{eq}}{\langle 1/\tau_i\rangle_{eq}}
\end{align}
and the expression for the diffusion constant becomes
\begin{align}
\label{dpred}
D_{MB}&\approx \frac{\Delta}{6M}\langle(1- 2p_i^2)/\tau_i\rangle_{eq}
\end{align}

\begin{table}
\caption{The branching ratio, $w_{ji}\tau_i$ of four different metabasins at $T=0.52$}
\label{TABbranch}
\begin{tabular}{c|cccc}
\hline\hline
i & A & B & C & D \\\hline
$w_{ji}\tau_i$ 
&0.87  & 0.61 & 0.11& 0.13\\
&0.05 & 0.24   & 0.06 & 0.07\\
&0.02 & 0.06  & 0.05 & 0.05\\
&0.007 & 0.01  & 0.04 & 0.04\\
&0.006 & 0.01  & 0.04 & 0.03\\
&0.003 & 0.01  & 0.04 & 0.03\\
&... & ... & ... & ...\\
\hline
$p_i^2$ & 0.76 & 0.43& 0.03 &0.04 \\
\hline\hline
\end{tabular}
\end{table}

The branching ratios $w_{ji}\tau_i$ are measured for all the sampled metabasins, and $p^2_i$s
are calculated.
As shown by the examples in Figure~\ref{p2_T},
the return probability $p^2_i$ is relatively low at high temperatures.
But, as the temperature decreases, $p^2_i$ increases significantly for some metabasins.
This is because some of the transition branches are suppressed as temperature decreases.
According to Eq.~\eqref{dpred}, high $p^2_i$ makes the diffusion constant smaller.
Increasing values of  $p^2_i$ is , therefore, another source of the slow dynamics, and one that is not captured in the
trap model.

In Figure~\ref{diffusionconstant}, the prediction of Eq.~\eqref{dpred} is compared to measured $D(T)$.
It shows that incorporating the return probability $p^2_i$ has a significant effect on $D_{MB}$  and improves the
agreement with the measured $D$.
In \cite{2003PhRvE..67c0501D}, Doliwa and Heuer were able to predict the diffusion constant
without considering the return probability,
due to the fact that the metabasins were constructed from trajectories, and the inherent structures that
were connected by quick hops were grouped together into the same metabasin.  
The dynamics in the space of MB defined in this manner satisfies more stringent requirements\cite{rubner:011504}, inducing $p_i^2=0$.
In this work, metabasins were instead constructed from the transition branch ratio of inherent structures, without any
consideration of the absolute values of the barrier heights between metabasins.   Our construction gives rise to a
graph which is much more inhomogeneous and the mean-field assumption of the trap model is not a good approximation.
The advantage of this method of construction is that it gives a better idea of the connectivity.   To emphasize, the
effective dynamics in the metabasin space is certainly Markovian but the random walk lives on a graph with non-trivial
connectivity.

\figone{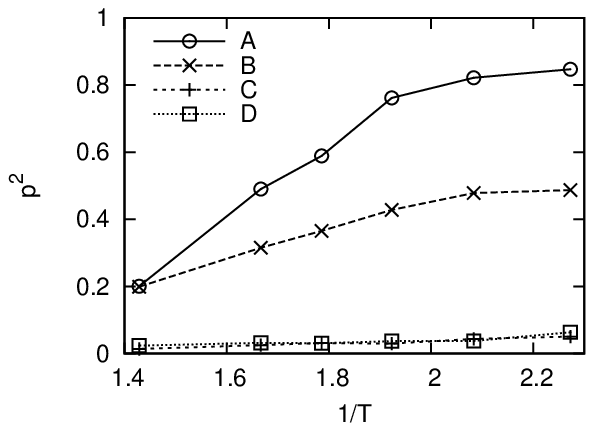}{Comparison of $p^2_i(T)$, the return probabilities of the four metabasins listed in Table~\ref{TABbranch}}{p2_T}

\figone{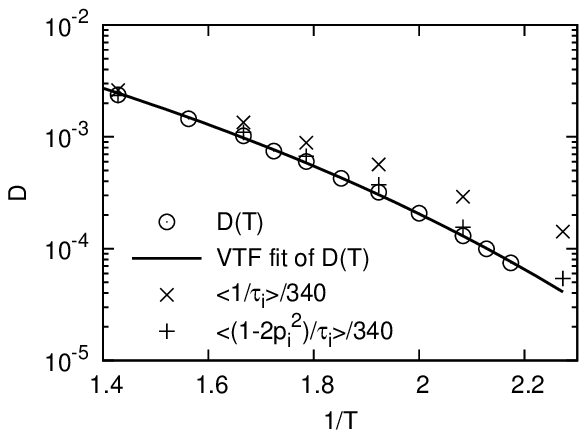}{Comparison of the temperature dependence of the measured diffusion constant, $D,$ of type $A$
particles, with the predictions of the trap model, and the trap model modified to include a finite return probability.
The relationship between the diffusion constant and $\tau_{i}$ in Eq. \ref{dpred} has an undetermined constant, which
was chosen to be $1/340$ in this figure.}{diffusionconstant}
\section{Summary}
In this work, we have presented a technique for constructing metabasins using branching ratios of transitions  between
inherent structures.  The dynamics in the space of these metabasins is Markovian, and can be described as an activated
process in an energy landscape.  A mean field trap model that ignores the connectivity
of the metabasins, however, underestimates the slowing down of the dynamics with decreasing temperature. 
We show that incorporating the connectivity through a set of return probabilities leads to quantitative agreement
between the predicted and measured temperature dependence of the diffusion constant.
This result demonstrates that connectivity is an essential element of the effective dynamics in the 
metabasin space.
At temperatures higher than $\simeq 0.7$, the mean waiting time of metabasins
approachs the ballistic time scale, and mapping the dynamics to metabasin space loses its utility. At temperatures
lower than 
$\simeq 0.45$, the return probabilities become so large that higher order terms in Eq.\eqref{msdfull}
need to be included and the effective dynamics loses its simplicity.   The mapping to metabasin space is, therefore,
most useful in the range $0.45 < T < 0.7$
At the high end of
this temperature range,  the connectivity is reasonably
homogeneous, and the return probabilities are low.  In this regime, effective trap models capture the essential
features of the slow dynamics. 
As the system is cooled, some of the connections are suppressed more strongly then
the others.  The connectivity becomes sparse, resulting in higher return probabilities, and break down of the
assumption, underlying the trap model, that once the system escapes fro a trap it chooses one from a fixed
distribution of trap depths that does not have any connectivity information. 
It is clear that the difference between connections, i.e., the inhomogeniety of connectivity,
causes the connectivity to be sparse at low temperatures.
At the same time, we find no obvious correlation between return probability and metabasin
energy, and therefore, metabasins of different energies seems to have similiar connectivity environment.

Instead of metabasins, J. Kim and T. Keyes\cite{2004kim} have considered the strong return dynamics between
inherent structures of a supercooled CS\subscript{2} system, and postulated that the
return dynamics could be coarse-grained to the motion within metabasins.
In this paper, however, we showed that even after coarse-graining
the strongly connected inherent structures into metabasins at a temperature in the middle of the range of interest to
the slow dynamics, 
the connectivity between metabasins plays a significant rule,
especially at low temperatures.   Return dynamics appears to be  less important for some small
systems\cite{PhysRevE.64.032201}.  In general, however, if the multiple funnel landscape\cite{doye-1999-111,
d.wales-2000} is an adequate representation of the energy minima, then any definition of metabasins based on a single
temperature will have to include the influence of return dynamics at low enough temperatures.

We acknowledge helpful discussions with David Reichman, especially during the early stages of this work. This work was
supported by NSF DMR-0549762.

\bibliography{mpaper}

\begin{thebibliography}{29}
\expandafter\ifx\csname natexlab\endcsname\relax\def\natexlab#1{#1}\fi
\expandafter\ifx\csname bibnamefont\endcsname\relax
  \def\bibnamefont#1{#1}\fi
\expandafter\ifx\csname bibfnamefont\endcsname\relax
  \def\bibfnamefont#1{#1}\fi
\expandafter\ifx\csname citenamefont\endcsname\relax
  \def\citenamefont#1{#1}\fi
\expandafter\ifx\csname url\endcsname\relax
  \def\url#1{\texttt{#1}}\fi
\expandafter\ifx\csname urlprefix\endcsname\relax\def\urlprefix{URL }\fi
\providecommand{\bibinfo}[2]{#2}
\providecommand{\eprint}[2][]{\url{#2}}

\bibitem[{\citenamefont{Goldstein}(1969)}]{Goldstein1969}
\bibinfo{author}{\bibfnamefont{M.}~\bibnamefont{Goldstein}},
  \bibinfo{journal}{J.\ Chem.\ Phys.} \textbf{\bibinfo{volume}{51}},
  \bibinfo{pages}{3728} (\bibinfo{year}{1969}).

\bibitem[{\citenamefont{Stillinger and Weber}(1984)}]{StillingerWeber1984}
\bibinfo{author}{\bibfnamefont{F.~H.} \bibnamefont{Stillinger}}
  \bibnamefont{and} \bibinfo{author}{\bibfnamefont{T.~A.} \bibnamefont{Weber}},
  \bibinfo{journal}{Science} \textbf{\bibinfo{volume}{225}},
  \bibinfo{pages}{983} (\bibinfo{year}{1984}).

\bibitem[{\citenamefont{Stillinger}(1995)}]{Stillinger1995}
\bibinfo{author}{\bibfnamefont{F.~H.} \bibnamefont{Stillinger}},
  \bibinfo{journal}{Science} \textbf{\bibinfo{volume}{267}},
  \bibinfo{pages}{1935} (\bibinfo{year}{1995}).

\bibitem[{\citenamefont{B\"uchner and Heuer}(1999)}]{PhysRevE.60.6507}
\bibinfo{author}{\bibfnamefont{S.}~\bibnamefont{B\"uchner}} \bibnamefont{and}
  \bibinfo{author}{\bibfnamefont{A.}~\bibnamefont{Heuer}},
  \bibinfo{journal}{Phys. Rev. E} \textbf{\bibinfo{volume}{60}},
  \bibinfo{pages}{6507} (\bibinfo{year}{1999}).

\bibitem[{\citenamefont{{Schr{\o}der} et~al.}(2000)\citenamefont{{Schr{\o}der},
  {Sastry}, {Dyre}, and {Glotzer}}}]{2000JChPh.112.9834S}
\bibinfo{author}{\bibfnamefont{T.~B.} \bibnamefont{{Schr{\o}der}}},
  \bibinfo{author}{\bibfnamefont{S.}~\bibnamefont{{Sastry}}},
  \bibinfo{author}{\bibfnamefont{J.~C.} \bibnamefont{{Dyre}}},
  \bibnamefont{and} \bibinfo{author}{\bibfnamefont{S.~C.}
  \bibnamefont{{Glotzer}}}, \bibinfo{journal}{\jcp}
  \textbf{\bibinfo{volume}{112}}, \bibinfo{pages}{9834} (\bibinfo{year}{2000}),
  \eprint{cond-mat/9901271}.

\bibitem[{\citenamefont{Heuer}(2008)}]{Heuer0953-8984-20-37-373101}
\bibinfo{author}{\bibfnamefont{A.}~\bibnamefont{Heuer}},
  \bibinfo{journal}{Journal of Physics: Condensed Matter}
  \textbf{\bibinfo{volume}{20}}, \bibinfo{pages}{373101}
  (\bibinfo{year}{2008}).

\bibitem[{\citenamefont{B\"uchner and Heuer}(2000)}]{PhysRevLett.84.2168}
\bibinfo{author}{\bibfnamefont{S.}~\bibnamefont{B\"uchner}} \bibnamefont{and}
  \bibinfo{author}{\bibfnamefont{A.}~\bibnamefont{Heuer}},
  \bibinfo{journal}{Phys. Rev. Lett.} \textbf{\bibinfo{volume}{84}},
  \bibinfo{pages}{2168} (\bibinfo{year}{2000}).

\bibitem[{\citenamefont{Doliwa and Heuer}(2003)}]{PhysRevLett.91.235501}
\bibinfo{author}{\bibfnamefont{B.}~\bibnamefont{Doliwa}} \bibnamefont{and}
  \bibinfo{author}{\bibfnamefont{A.}~\bibnamefont{Heuer}},
  \bibinfo{journal}{Phys. Rev. Lett.} \textbf{\bibinfo{volume}{91}},
  \bibinfo{pages}{235501} (\bibinfo{year}{2003}).

\bibitem[{\citenamefont{{Doliwa} and {Heuer}}(2003)}]{2003PhRvE..67c0501D}
\bibinfo{author}{\bibfnamefont{B.}~\bibnamefont{{Doliwa}}} \bibnamefont{and}
  \bibinfo{author}{\bibfnamefont{A.}~\bibnamefont{{Heuer}}},
  \bibinfo{journal}{\pre} \textbf{\bibinfo{volume}{67}},
  \bibinfo{pages}{030501} (\bibinfo{year}{2003}), \eprint{cond-mat/0205283}.

\bibitem[{\citenamefont{{Monthus} and {Bouchaud}}(1996)}]{1996JPhA...29.3847M}
\bibinfo{author}{\bibfnamefont{C.}~\bibnamefont{{Monthus}}} \bibnamefont{and}
  \bibinfo{author}{\bibfnamefont{J.-P.} \bibnamefont{{Bouchaud}}},
  \bibinfo{journal}{Journal of Physics A Mathematical General}
  \textbf{\bibinfo{volume}{29}}, \bibinfo{pages}{3847} (\bibinfo{year}{1996}),
  \eprint{cond-mat/9601012}.

\bibitem[{\citenamefont{Denny et~al.}(2003)\citenamefont{Denny, Reichman, and
  Bouchaud}}]{PhysRevLett.90.025503}
\bibinfo{author}{\bibfnamefont{R.~A.} \bibnamefont{Denny}},
  \bibinfo{author}{\bibfnamefont{D.~R.} \bibnamefont{Reichman}},
  \bibnamefont{and} \bibinfo{author}{\bibfnamefont{J.-P.}
  \bibnamefont{Bouchaud}}, \bibinfo{journal}{Phys. Rev. Lett.}
  \textbf{\bibinfo{volume}{90}}, \bibinfo{pages}{025503}
  (\bibinfo{year}{2003}).

\bibitem[{\citenamefont{{Heuer} et~al.}(2005)\citenamefont{{Heuer}, {Doliwa},
  and {Saksaengwijit}}}]{2005PhRvE..72b1503H}
\bibinfo{author}{\bibfnamefont{A.}~\bibnamefont{{Heuer}}},
  \bibinfo{author}{\bibfnamefont{B.}~\bibnamefont{{Doliwa}}}, \bibnamefont{and}
  \bibinfo{author}{\bibfnamefont{A.}~\bibnamefont{{Saksaengwijit}}},
  \bibinfo{journal}{\pre} \textbf{\bibinfo{volume}{72}},
  \bibinfo{pages}{021503} (\bibinfo{year}{2005}), \eprint{cond-mat/0503211}.

\bibitem[{\citenamefont{Doye et~al.}(1999)\citenamefont{Doye, Miller, and
  Wales}}]{doye-1999-111}
\bibinfo{author}{\bibfnamefont{J.}~\bibnamefont{Doye}},
  \bibinfo{author}{\bibfnamefont{M.}~\bibnamefont{Miller}}, \bibnamefont{and}
  \bibinfo{author}{\bibfnamefont{D.}~\bibnamefont{Wales}}, \bibinfo{journal}{J.
  Chem. Phys.} \textbf{\bibinfo{volume}{111}}, \bibinfo{pages}{8417}
  (\bibinfo{year}{1999}).

\bibitem[{\citenamefont{Wales et~al.}(2000)\citenamefont{Wales, Doye, Miller,
  Mortenson, and Walsh}}]{d.wales-2000}
\bibinfo{author}{\bibfnamefont{D.}~\bibnamefont{Wales}},
  \bibinfo{author}{\bibfnamefont{J.}~\bibnamefont{Doye}},
  \bibinfo{author}{\bibfnamefont{M.}~\bibnamefont{Miller}},
  \bibinfo{author}{\bibfnamefont{P.}~\bibnamefont{Mortenson}},
  \bibnamefont{and} \bibinfo{author}{\bibfnamefont{T.}~\bibnamefont{Walsh}},
  \bibinfo{journal}{Advances in Chemical Physics}
  \textbf{\bibinfo{volume}{115}}, \bibinfo{pages}{1} (\bibinfo{year}{2000}).

\bibitem[{\citenamefont{Kob and Andersen}(1994)}]{PhysRevLett.73.1376}
\bibinfo{author}{\bibfnamefont{W.}~\bibnamefont{Kob}} \bibnamefont{and}
  \bibinfo{author}{\bibfnamefont{H.~C.} \bibnamefont{Andersen}},
  \bibinfo{journal}{Phys. Rev. Lett.} \textbf{\bibinfo{volume}{73}},
  \bibinfo{pages}{1376} (\bibinfo{year}{1994}).

\bibitem[{\citenamefont{Allen and Tildesley}(1987)}]{AllenTildesley}
\bibinfo{author}{\bibfnamefont{M.~P.} \bibnamefont{Allen}} \bibnamefont{and}
  \bibinfo{author}{\bibfnamefont{D.~J.} \bibnamefont{Tildesley}},
  \emph{\bibinfo{title}{Computer simulation of liquids}}
  (\bibinfo{publisher}{Clarendon Press}, \bibinfo{address}{New York, NY, USA},
  \bibinfo{year}{1987}), ISBN \bibinfo{isbn}{0-19-855375-7}.

\bibitem[{\citenamefont{Stoddard and Ford}(1973)}]{PhysRevA.8.1504}
\bibinfo{author}{\bibfnamefont{S.~D.} \bibnamefont{Stoddard}} \bibnamefont{and}
  \bibinfo{author}{\bibfnamefont{J.}~\bibnamefont{Ford}},
  \bibinfo{journal}{Phys. Rev. A} \textbf{\bibinfo{volume}{8}},
  \bibinfo{pages}{1504} (\bibinfo{year}{1973}).

\bibitem[{\citenamefont{{S} and {Sastry}}(2003)}]{2003JPCM...15.1253S}
\bibinfo{author}{\bibfnamefont{A.~S.} \bibnamefont{{S}}} \bibnamefont{and}
  \bibinfo{author}{\bibfnamefont{S.}~\bibnamefont{{Sastry}}},
  \bibinfo{journal}{Journal of Physics Condensed Matter}
  \textbf{\bibinfo{volume}{15}}, \bibinfo{pages}{1253} (\bibinfo{year}{2003}).

\bibitem[{\citenamefont{{Coslovich} and {Pastore}}(2007)}]{2007JChPh.127l4504C}
\bibinfo{author}{\bibfnamefont{D.}~\bibnamefont{{Coslovich}}} \bibnamefont{and}
  \bibinfo{author}{\bibfnamefont{G.}~\bibnamefont{{Pastore}}},
  \bibinfo{journal}{\jcp} \textbf{\bibinfo{volume}{127}},
  \bibinfo{pages}{124504} (\bibinfo{year}{2007}), \eprint{0705.0626}.

\bibitem[{\citenamefont{{Sastry} et~al.}(1998)\citenamefont{{Sastry},
  {Debenedetti}, and {Stillinger}}}]{1998Natur.393..554S}
\bibinfo{author}{\bibfnamefont{S.}~\bibnamefont{{Sastry}}},
  \bibinfo{author}{\bibfnamefont{P.~G.} \bibnamefont{{Debenedetti}}},
  \bibnamefont{and} \bibinfo{author}{\bibfnamefont{F.~H.}
  \bibnamefont{{Stillinger}}}, \bibinfo{journal}{\nat}
  \textbf{\bibinfo{volume}{393}}, \bibinfo{pages}{554} (\bibinfo{year}{1998}).

\bibitem[{\citenamefont{Mauro et~al.}(2007)\citenamefont{Mauro, Gupta, and
  Loucks}}]{mauro:184511}
\bibinfo{author}{\bibfnamefont{J.~C.} \bibnamefont{Mauro}},
  \bibinfo{author}{\bibfnamefont{P.~K.} \bibnamefont{Gupta}}, \bibnamefont{and}
  \bibinfo{author}{\bibfnamefont{R.~J.} \bibnamefont{Loucks}},
  \bibinfo{journal}{The Journal of Chemical Physics}
  \textbf{\bibinfo{volume}{126}}, \bibinfo{eid}{184511}
  (pages~\bibinfo{numpages}{11}) (\bibinfo{year}{2007}),
  \urlprefix\url{http://link.aip.org/link/?JCP/126/184511/1}.

\bibitem[{\citenamefont{Kampen}(1992)}]{Van-Kampen}
\bibinfo{author}{\bibfnamefont{N.~G.~V.} \bibnamefont{Kampen}},
  \emph{\bibinfo{title}{Stochastic processes in physics and chemistry}}
  (\bibinfo{publisher}{North-Holland}, \bibinfo{address}{Amsterdam},
  \bibinfo{year}{1992}).

\bibitem[{\citenamefont{Sciortino et~al.}(1999)\citenamefont{Sciortino, Kob,
  and Tartaglia}}]{PhysRevLett.83.3214}
\bibinfo{author}{\bibfnamefont{F.}~\bibnamefont{Sciortino}},
  \bibinfo{author}{\bibfnamefont{W.}~\bibnamefont{Kob}}, \bibnamefont{and}
  \bibinfo{author}{\bibfnamefont{P.}~\bibnamefont{Tartaglia}},
  \bibinfo{journal}{Phys. Rev. Lett.} \textbf{\bibinfo{volume}{83}},
  \bibinfo{pages}{3214} (\bibinfo{year}{1999}).

\bibitem[{\citenamefont{Hansen and McDonald}(1986)}]{hansen86a}
\bibinfo{author}{\bibfnamefont{J.~P.} \bibnamefont{Hansen}} \bibnamefont{and}
  \bibinfo{author}{\bibfnamefont{I.~R.} \bibnamefont{McDonald}},
  \emph{\bibinfo{title}{Theory of Simple Liquids}}
  (\bibinfo{publisher}{Academic Press}, \bibinfo{address}{London},
  \bibinfo{year}{1986}).

\bibitem[{\citenamefont{{Szamel} and {Flenner}}(2006)}]{2006PhRvE..73a1504S}
\bibinfo{author}{\bibfnamefont{G.}~\bibnamefont{{Szamel}}} \bibnamefont{and}
  \bibinfo{author}{\bibfnamefont{E.}~\bibnamefont{{Flenner}}},
  \bibinfo{journal}{\pre} \textbf{\bibinfo{volume}{73}},
  \bibinfo{pages}{011504} (\bibinfo{year}{2006}),
  \eprint{arXiv:cond-mat/0508108}.

\bibitem[{\citenamefont{{Schulz} et~al.}(2001)\citenamefont{{Schulz},
  {Trimper}, and {Schulz}}}]{2001JChPh.11410402S}
\bibinfo{author}{\bibfnamefont{B.~M.} \bibnamefont{{Schulz}}},
  \bibinfo{author}{\bibfnamefont{S.}~\bibnamefont{{Trimper}}},
  \bibnamefont{and} \bibinfo{author}{\bibfnamefont{M.}~\bibnamefont{{Schulz}}},
  \bibinfo{journal}{\jcp} \textbf{\bibinfo{volume}{114}},
  \bibinfo{pages}{10402} (\bibinfo{year}{2001}).

\bibitem[{\citenamefont{Rubner and Heuer}(2008)}]{rubner:011504}
\bibinfo{author}{\bibfnamefont{O.}~\bibnamefont{Rubner}} \bibnamefont{and}
  \bibinfo{author}{\bibfnamefont{A.}~\bibnamefont{Heuer}},
  \bibinfo{journal}{Physical Review E (Statistical, Nonlinear, and Soft Matter
  Physics)} \textbf{\bibinfo{volume}{78}}, \bibinfo{eid}{011504}
  (pages~\bibinfo{numpages}{4}) (\bibinfo{year}{2008}),
  \urlprefix\url{http://link.aps.org/abstract/PRE/v78/e011504}.

\bibitem[{\citenamefont{Kim and Keyes}(2004)}]{2004kim}
\bibinfo{author}{\bibfnamefont{J.}~\bibnamefont{Kim}} \bibnamefont{and}
  \bibinfo{author}{\bibfnamefont{T.}~\bibnamefont{Keyes}},
  \bibinfo{journal}{The Journal of Chemical Physics} pp.
  \bibinfo{pages}{4237--4245} (\bibinfo{year}{2004}).

\bibitem[{\citenamefont{Keyes and Chowdhary}(2001)}]{PhysRevE.64.032201}
\bibinfo{author}{\bibfnamefont{T.}~\bibnamefont{Keyes}} \bibnamefont{and}
  \bibinfo{author}{\bibfnamefont{J.}~\bibnamefont{Chowdhary}},
  \bibinfo{journal}{Phys. Rev. E} \textbf{\bibinfo{volume}{64}},
  \bibinfo{pages}{032201} (\bibinfo{year}{2001}).

\end{thebibliography}
\end{document}